# ABOUT THE FIRST LAYER EFFECT IN SURFACE ENHANCED SPECTROSCOPY

[1]V.P. Chelibanov, [2]A.M. Polubotko

[1]State University of Information Technologies, Mechanics and Optics, Kronverkskii 49, 197101 Saint Petersburg, RUSSIA  E-mail: Chelibanov@gmail.com

[2] A.F. Ioffe Physico-Technical Institute, Politechnicheskaya 26, 194021 Saint Petersburg, RUSSIA,Tel: (812) 274-77-29, Fax: (812) 297-10-17, E-mail: alex.marina@mail.ioffe.ru

The first layer effect in Surface enhanced spectroscopy (SES) is a well known phenomenon. Its essence is that the molecules, adsorbed in the first layer enhance SERS, SEHRS and the infrared absorption significantly larger than the molecules, adsorbed in the second and upper layers [1,2]. Usually researchers associate this additional enhancement with contact of the molecules with the substrate and with the change of the molecules wavefunctions due to this contact. This effect is called as a chemical enhancement also. This opinion raises many investigations. However from our point of view it is a mistake and the nature of this effect is associated not with chemistry, but with a very strong change of the electric field and its derivatives, when one move from the surface. In the Dipole-Quadrupole theory of all the above mentioned processes [3-5], which we use here, the enhancement arises due to the strong enhancement of the dipole interaction, associated with large increase of the electric field and the enhancement of the quadrupole interaction due to strong increase of the field derivatives with the same indices $\frac{\partial E_i}{\partial x_i}$. The electric field and its derivatives strongly differ in the second and upper layers of the adsorbed molecules, compared with the ones, in the first layer, that causes this strong difference in the enhancement.

As it is well accepted in Surface Enhanced Spectroscopy, the main enhancement in SES arises in the so called active sites, or hot spots. One type of the active sites are the tops of such roughness as a wedges, or cones, or spikes, where the electric field behavior is singular. As it is well known this behavior can be described by the formula

$$E_r \sim |\overline{E}_{o,inc}| C_0 \left(\frac{l}{r}\right)^\beta \qquad (1)$$

where $|\overline{E}_{o,inc}|$ is the amplitude of the incident electric field, $C_0 \sim 1$ is a numerical coefficient, $l$ is a characteristic size of the roughness, $r$ - is the distance from the top of the roughness, $0 < \beta < 1$ and depends on the angle near the top of the roughness (Figure 1).

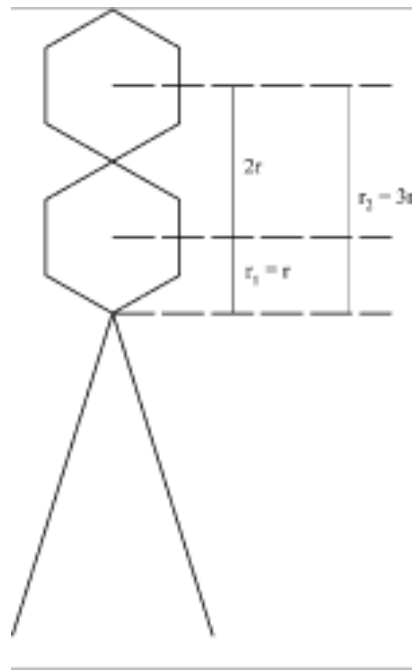

Figure 1. Molecules adsorbed on the top of the model roughness of the wedge, cone or spike form in the first and the second layer.

Let us consider the relation of the electric field and its derivatives in the first and the second layer for SERS, SEHRS and SEIRA. In accordance with Figure 1 these relations are

$$\left(\frac{r_2}{r_1}\right)^{2n\beta} = (3)^{2n\beta} \tag{2}$$

for the pure dipole mechanism of the enhancement and

$$\left(\frac{r_2}{r_1}\right)^{2n+2n\beta} = (3)^{2n+2n\beta} \tag{3}$$

for the pure quadrupole mechanism. Here $n = 2,3,1$ for these processes and is the order of the processes. One can see, that for the pure dipole mechanism we can easily obtain the relative enhancements in the first layer ~ 80, 700 and 9 respectively (here we put $\beta = 1$) and for the pure quadrupole mechanism the relative enhancements ~ 6400, 50000, 80. Taking into account that we deal with some ideal situation one can understand that in a real situation with a more real roughness (Figure 2), these relative enhancement coefficients can be significantly lower.

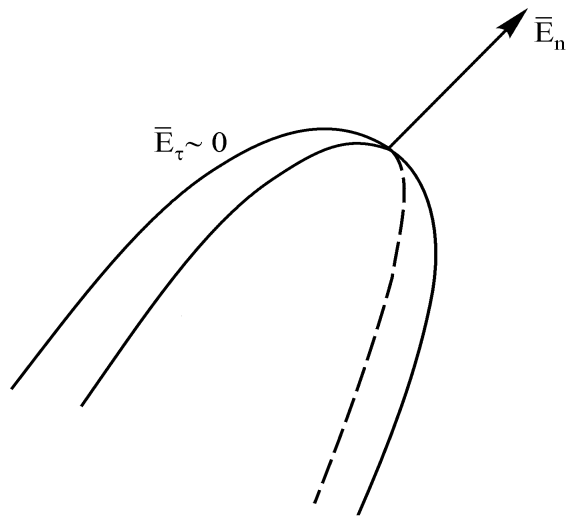

Figure 2. A real roughness.

However the obtained relative enhancements are so strong that it is obvious that the first layer effect is easily can be explained by the pure electrodynamical mechanism in the Dipole-

Quadrupole theory and is associated with very strong change of the electric field and its derivatives, when one moves away from the surface.